\documentclass[11pt]{article}
\usepackage{epsf}
\usepackage{latexsym}

\newcommand{\ap}[3]{{\sl Ann.~Phys.} {\bf #1} (19#2) #3}

\newcommand{\np}[3]{{\sl Nucl. Phys.} {\bf #1} (19#2)~#3}
\newcommand{\pl}[3]{{\sl Phys. Lett.} {\bf #1} (19#2) #3}
\newcommand{\pr}[3]{{\sl Phys. Rev.} {\bf #1} (19#2) #3}

\newcommand{\vj}[4]{{\sl #1~}{\bf #2} (19#3) #4}

\def\be{\begin{equation}}
\def\ee{\end{equation}}
\def\bea{\begin{eqnarray}}
\def\eea{\end{eqnarray}}

\makeatletter 
\def\secteqno{\@addtoreset{equation}{section}%
\def\theequation{\thesection.\arabic{equation}}} 
 
\def\endsecteqno{\def{theequation\{\@ifundefined{chapter}%
{\arabic{equation}}{\thechapter.\arabic{equation}}}} 
\makeatother

\newcommand{\bfi}[1]{\begin{figure}[#1]}
\newcommand{\efi}{\end{figure}}
\newcommand{\bpi}[2]{\begin{picture}(#1,#2)}
\newcommand{\epi}{\end{picture}}

\newcommand{\nn}{\nonumber\\}

\def\ie{{\it i.e.\/}}
\def\eg{e.g.}

\newcommand{\prop}{\Delta}

\newcommand{\dsl}{{\not \! \partial}}
\renewcommand{\d}{\partial}
\newcommand{\ii}{{\mathrm i}}
\newcommand{\T}{{\mathrm T}}
\newcommand{\Q}{{\mathrm Q}}
\newcommand{\B}{{\mathrm B}}
\newcommand{\A}{{\mathrm A}}

\newcommand{\OO}{{\mathcal O}}
\newcommand{\al}{\alpha}

\begin{document}

\pagestyle{empty} 
{\hfill \parbox{6cm}{\begin{center} 
	UG-FT-89/98  \\ 
	hep-ph/9808071 \\
	August 1998                   
\end{center}}} 
	      
\vspace*{1cm}                               
\begin{center} 
\large{\bf Constrained differential renormalization \\ }
\large{\bf of Yang-Mills theories} 
\vskip .6truein 
\centerline {\large 
M. P\'erez-Victoria}  
\end{center} 
\vspace{.3cm} 
\begin{center}
{Dpto. de F\'{\i}sica Te\'orica y del Cosmos,  
 Universidad de Granada, \\
 18071 Granada, Spain} 
\end{center}
\vspace{1.5cm} 
 
\centerline{\bf Abstract} 
\medskip 

We renormalize QCD to one loop in coordinate space using constrained 
differential renormalization, and show explicitly that the Slavnov-Taylor 
identities are preserved by this method.

\vspace*{2cm}

\pagestyle{plain}

Differential regularization and renormalization \cite{FJL} is a method 
that works directly on Feynman graphs in coordinate space, substituting 
singular expressions by derivatives of well-behaved distributions. It has
proved to be quite simple and convenient in a number of applications.  
In gauge theories, however, a problematic feature arises: the Ward 
identities must be studied explicitly to fix the arbitrariness of the method 
in such a way that gauge invariance is preserved.

A solution at the one-loop level is the symmetric procedure of differential
renormalization proposed in Ref.~\cite{CDR}. This so-called constrained
differential renormalization (CDR) introduces no ambiguities and has been shown
to preserve Abelian gauge invariance \cite{CDR,techniques} and supersymmetry in
a non-trivial calculation in supergravity \cite{g2}. 
It is the purpose of this letter to
apply CDR to a non-Abelian gauge theory and study the corresponding
Slavnov-Taylor identities \cite{ST}. We shall consider a Yang-Mills 
theory with gauge group $SU(N_c)$ coupled to $N_f$ Dirac fermions in the 
fundamental representation, \ie, QCD with 
$N_c$ colours and $N_f$ quark flavours. 
The calculation of the gluon selfenergy and the triple gluon vertex, using
conventional differential renormalization, was carried out 
in Refs. \cite{FJL} and \cite{FGJR}, respectively. The background field
method \cite{BGM} was employed there because it allows a much more direct
determination of the $\beta$ function and leads to simpler Ward identities.
We use the conventional formalism instead, precisely for the last reason: we
want to test CDR in the most complex case, and show that it preserves the
(more involved) Slavnov-Taylor identities rather than the Ward-Takahashi
like identities of the background field formalism. Nevertheless, for
comparison with Refs. \cite{FJL,FGJR}, we have also
applied CDR to the background field calculations mentioned above.

In Ref. \cite{techniques} it was argued that CDR preserves the Ward identities
of Abelian gauge invariance because it maintains the properties that are 
required for their derivation, like the fulfilment of equations of motion for
renormalized expressions or the commutativity of differentiation with 
renormalization. 
Actually, this argument applies equally well to the 
case of non-Abelian gauge invariance, since the structure of the 
interaction Lagrangian is never used. 
Of course, the symmetry of the Lagrangian is 
essential for the fulfilment of the corresponding Ward identities, but this 
is a matter of {\em combinatorics} \cite{Diagrammar} and CDR does not 
interfere
with it (essentially, CDR ensures that the building blocks of such
combinatorics behave correctly). A new feature of the non-Abelian case is
the appearance of composite operators in the Slavnov-Taylor identities 
(in the Zinn-Justin form \cite{ZJL}, which we shall use here). 
In Ref. \cite{FJL} it was shown that
differential renormalization can be directly applied to diagrams with operator
insertions. The same holds for the constrained procedure.

\begin{figure}[ht]
\setlength{\unitlength}{1cm}
\begin{center}
\begin{picture}(12,16)
\epsfxsize=8.5cm
\put(0,0){\epsfbox{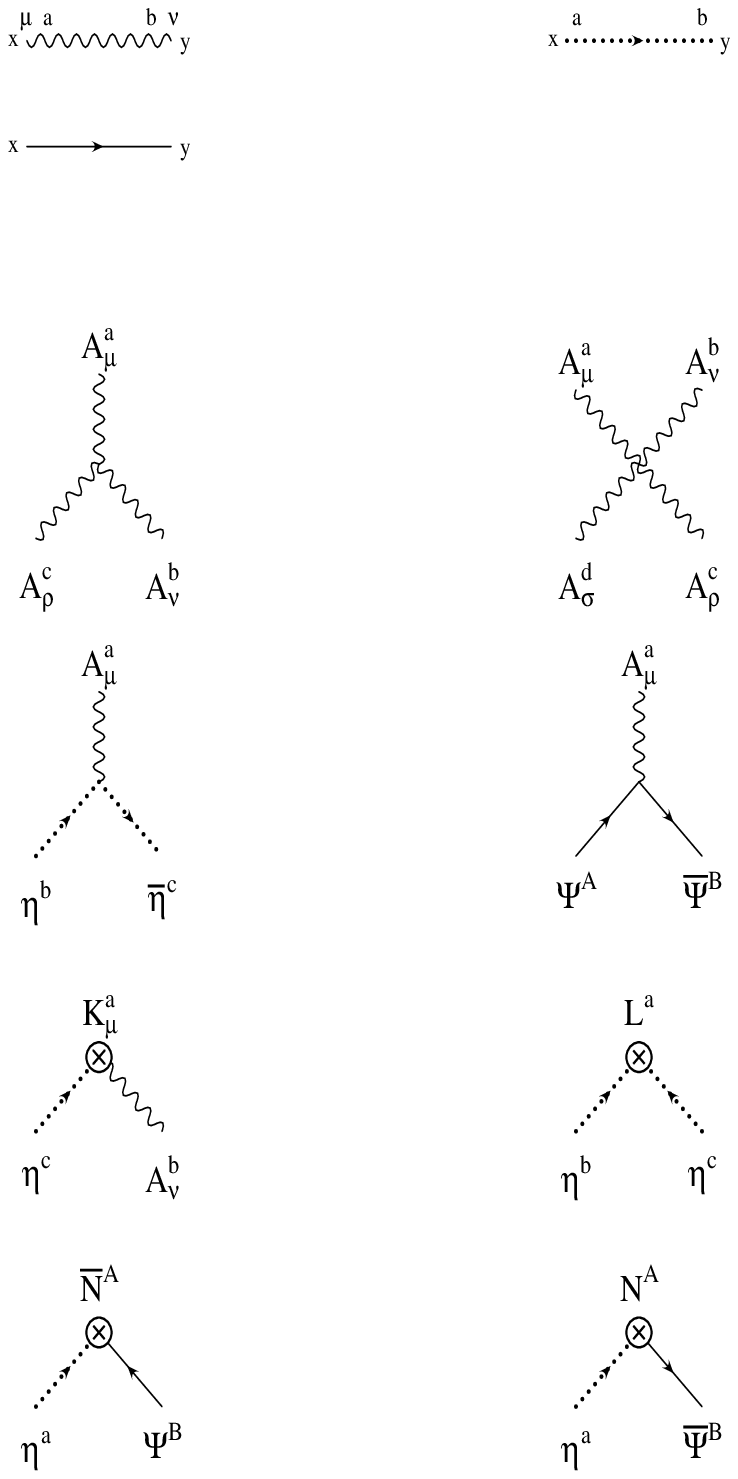}}
\put(2.5,16.4){$\delta_{\mu\nu}\delta^{ab}\prop(x-y)$}
\put(8.5,16.4){$\delta^{ab}\prop(x-y)$}
\put(2.5,15.2){$(\dsl^x + m) \prop_m(x-y)$}
\put(2.5,11.6){\parbox{5cm}{$g f^{abc} [\delta_{\mu\nu}
           (\d_\rho^{A_\mu^a}-\d_\rho^{A_\nu^b})$ \\
           $\mbox{} + \delta_{\nu\rho}(\d_\mu^{A_\nu^b}-\d_\mu^{A_\rho^c})$ \\
           $\mbox{} + \delta_{\rho\mu}(\d_\nu^{A_\rho^c}-\d_\nu^{A_\mu^a})$}}
\put(8.5,11.6){\parbox{5cm}{$-g^2[f^{abr}f^{cdr}
              (\delta_{\mu\rho}\delta_{\nu\sigma}-
              \delta_{\mu\sigma}\delta_{\nu\rho})$ \\
              $\mbox{} + f^{acr}f^{dbr}
              (\delta_{\mu\sigma}\delta_{\nu\rho}-
              \delta_{\mu\nu}\delta_{\rho\sigma}$ \\
              $\mbox{} + f^{adr}f^{bcr}
              (\delta_{\mu\nu}\delta_{\rho\sigma}-
              \delta_{\mu\rho}\delta_{\nu\sigma})]$}}
\put(2.5,8){$-g f^{abc} \d_\mu^{\bar{\eta}^c}$}
\put(8.5,8){$-g (T^a)^{BA} \gamma_\mu$}         
\put(2.5,4.6){$-g f^{abc} \delta_{\mu\nu}$}
\put(8.5,4.6){$-g f^{abc}$}
\put(2.5,1.4){$-g (T^a)^{AB}$}
\put(8.5,1.4){$-g (T^a)^{BA}$}
           
\end{picture}
\end{center}
\caption{Coordinate space Feynman rules for QCD, including 
insertions of BRST variations. In vertices, the derivatives
(with respect to the vertex space-time point) act on the
field indicated by the superscript. \label{fig_FR}}
\end{figure}

In the following, after writing the Lagrangian, 
we give the renormalized expressions in coordinate space of all the singular 
one-loop 1PI Green functions of elementary fields. The
diagrams with operator insertions that contribute to the Slavnov-Taylor
identities have also been calculated, but the explicit results are not given
here. Then, we write all the Slavnov-Taylor identities involving these
renormalized Green functions. We have used a symbolic computer program to 
verify that they are indeed fulfilled.
The QCD Lagrangian in the Feynman gauge, written in Euclidean space and
including ghost terms, reads
\be
  {\mathcal L} = \frac{1}{4} F^a_{\mu\nu} F^a_{\mu\nu} + 
  \frac{1}{2} (\d_\mu A^a_\mu) (\d_\nu A^a_\nu) + 
  \d_\mu \bar{\eta}^a (D_\mu \eta)^a + 
  \sum_{i=1}^{N_f} \bar{\Psi}_i ({\not \! D}+m_i) \Psi_i \, ,
\ee
with
\bea
  && F^a_{\mu\nu} =  \d_\mu A^a_\nu - \d_\nu A^a_\mu + 
  g f^{abc} A^b_\mu A^c_\nu \, ,  \\
  && (D_\mu \eta)^a  =  \d_\mu \eta^a + g f^{abc} A^b_\mu \eta^c \, ,  \\
  && D_\mu \Psi_i  =  \d_\mu \Psi_i - \ii g A^a_\mu T^a \Psi_i \, . 
\eea
$T^a$ are the $SU(N_c)$ generators in the fundamental
representation, and $f^{abc}$ the structure constants. This Lagrangian
is invariant under the BRST transformation \cite{BRST}
\bea
  \delta A_\mu^a &=& - (D_\mu \eta)^a \delta \lambda \equiv 
    s A_\mu^a \delta \lambda \, , \\
  \delta \bar{\eta}^a &=& - \d_\mu A_\mu^a \delta \lambda \equiv
    s \bar{\eta}^a \delta \lambda \, , \\
  \delta \eta^a &=& -\frac{g}{2} f^{abc} \eta^b \eta^c \delta \lambda
    \equiv s \eta^a \delta \lambda \, , \\
  \delta \Psi_i &=& -g T^a \eta^a \Psi_i \delta \lambda \equiv
    s \Psi_i \delta \lambda \, , \\
  \delta \bar{\Psi}_i &=& -g \bar{\Psi}_i T^a \eta^a \delta \lambda \equiv
    s \bar{\Psi}_i \delta \lambda \, ,
\eea
where $\delta \lambda$ is a constant Grassman parameter. In order to
obtain simple Slavnov-Taylor identities for the 1PI Green functions, it is
convenient to add to the Lagrangian source terms not only for elementary fields
but also for their BRST variations \cite{ZJL}:
\bea
  \lefteqn{{\mathcal L}_{\mbox{\footnotesize sources}}  = 
   - \left( J^a_\mu A^a_\mu +
  \bar{\eta}^a \xi^a + \bar{\xi}^a \eta^a + \sum_{i=1}^{N_f} 
  (\bar{\Psi}_i \chi_i + \bar{\chi}_i \Psi_i )  \right.} && \nn 
  && \mbox{} \left. + 
  K^a_\mu (sA^a_\mu) + L^a (s\eta^a) + \sum_{i=1}^{N_f} (
  \bar{N}_i (s\Psi_i) + (s\bar{\Psi}_i) N_i ) \right) \, .
\label{sources}
\eea
The sources of BRST variations are mere spectators in the Legendre transform
defining the generator functional of 1PI Green functions.
The coordinate space Feynman rules, including those for insertions of 
BRST variations, are displayed in Fig. \ref{fig_FR}. 

The singular 1PI Green functions of QCD at one loop are the gluon, quark and
ghost selfenergies, and the quark, ghost, triple gluon and quartic gluon 
vertices. The procedure to
calculate them with CDR is simple~\cite{CDR,techniques}. 
First, one expresses each contributing 
diagram in terms of the {\em basic functions} defined in Ref.~\cite{techniques}:
\bea
  && \A_m  \equiv  \prop_m(x) \delta(x) \, , \\
  && \B_{m_1m_2}[\OO](x)  \equiv  \prop_{m_1}(x) \OO^x \prop_{m_2}(x) \, , \\
  && \T_{m_1m_2m_3}[\OO](x,y)  \equiv  \prop_{m_1}(x) \prop_{m_2}(y)
    \OO^x \prop_{m_3}(x+y) \, , \\
  && \Q_{m_1m_2m_3m_4}[\OO](x,y,z)  \equiv  \prop_{m_1}(x) 
    \prop_{m_2}(y) \prop_{m_3}(z) \OO^x \prop_{m_4}(x+y+z) \, ,
\eea
where $\prop_m(x) = \frac{1}{4\pi^2} \frac{m K_1(mx)}{x}$ is the 
Feynman propagator in (Euclidean) coordinate space, with $K_1$ a modified
Bessel function, and $\OO$ is a differential operator. We shall suppress
mass subindices when all of them are zero.
Then, the singular basic functions are replaced by their renormalized 
expressions, which can be found in Tables 2, 3 and 4 of Ref.~\cite{techniques}.
For brevity, we refer to those tables and do not reproduce here the
renormalized basic functions that appear in our calculations. Also, since the
procedure of CDR has been explained in detail in Refs.~\cite{CDR,techniques},
we shall directly give the final result for each Green function.
The Feynman diagrams contributing to 1PI Green functions of elementary
fields can be found, \eg, in Ref.~\cite{PT}.

The renormalized gluon selfenergy in terms of renormalized basic functions
reads
\bea
  \lefteqn{<A_\mu^a(x_1)A_\nu^b(x_2)> = g^2 \delta^{ab} \left\{
  \rule[-.4cm]{0cm}{.8cm}
  N_c \left[ (3\d_\mu\d_\nu - 2\delta_{\mu\nu}\Box) \B^R[1]
  -4 \B^R[\d_\mu\d_\nu] \right] \right.} && \nn
  && \left. \mbox{} + 
  \sum_{i=1}^{N_f} \left[ \left(-2\d_\mu\d_\nu + \delta_{\mu\nu} 
  (\Box+2m_i^2) \right)
  \B_{m_im_i}^R[1] - 2 \B_{m_im_i}^R[\Box] + 4 \B_{m_im_i}^R[\d_\mu\d_\nu] 
  \right] \right\} \, ,
\label{gse}
\eea
where all functions, here and in the rest of two-point functions, 
depend on the coordinate difference $x=x_1-x_2$ and,
as prescribed by differential renormalization,
total derivatives are supposed to act formally by parts on
test functions \cite{FJL}. Throughout this paper,
$<\phi_1 \dots \phi_n>$ represents the one-loop correction to the
1PI Green function of the fields $\phi_1 \dots \phi_n$.
That the expression in Eq.~(\ref{gse}) is transverse, 
as required by gauge invariance,
is only apparent when the renormalized basic functions are substituted
by their explicit expressions in the mentioned tables. 
In the case of massless quarks the result is quite compact:
\bea
  \lefteqn{<A_\mu^a(x_1)A_\nu^b(x_2)>  = 
  - \frac{1}{144\pi^2} g^2 \delta^{ab} \,
  (\d_\mu\d_\nu-\delta_{\mu\nu} \Box)} && \nn
  && 
  \left[ (15N_c-6N_f) \frac{1}{4\pi^2} \Box
  \frac{\log x^2M^2}{x^2} + (2N_c-2N_f) \delta(x) \right] \, .
\eea
$M$ is the renormalization scale.
The dependence on the renormalization scale of Eq.~(\ref{gse}) is
\be
  M \frac{\d}{\d M}<A_\mu^a(x_1)A_\nu^b(x_2)>  =
  \frac{1}{24\pi^2} g^2 \delta^{ab} (5N_c-2N_f) 
  (\d_\mu\d_\nu - \delta_{\mu\nu} \Box) \delta(x) \, .
\ee
In the background field method, the background gluon selfenergy renormalized
with CDR reads (for massless quarks) 
\bea
  \lefteqn{<B_\mu^a(x_1)B_\nu^b(x_2)> =
   - \frac{1}{144\pi^2} g^2 \delta^{ab} \,
  (\d_\mu\d_\nu-\delta_{\mu\nu} \Box)} && \nn
  && 
  \left[ (33N_c-6N_f) \frac{1}{4\pi^2} \Box
  \frac{\log x^2M^2}{x^2} + (2N_c-2N_f) \delta(x) \right] \, ,
\label{BGgse}
\eea
and differs from the result (for $N_f=0$) in Ref.~\cite{FJL} by a finite 
local term (which can be absorbed into a redefinition of $M$). The one-loop
$\beta$ function of QCD, $\beta=-\frac{g^3}{48\pi^2}(11N_c-2N_f)$, 
can be directly read from the scale dependent part of Eq.~(\ref{BGgse}).

The {\bf quark selfenergy} is proportional to
the corresponding Green function of QED. 
In terms of basic functions, the 
renormalized quark selfenergy reads
\bea
  <\Psi_i^A(x_1)\bar{\Psi}_i^B(x_2)>  & = & g^2 \delta^{AB} \frac{N_c^2-1}{N_c}
  \left(\B^R_{0m_i}[\dsl] - 2m_i \B^R_{0m_i}[1] \right)  \nn
  &=&
  \frac{1}{64\pi^4} g^2 \delta^{AB} \frac{N_c^2-1}{N_c}
  \left\{ (\dsl-4m_i) \left[(\Box-m_i^2)
  \frac{K_0(m_ix)}{x} \right. \right. \nn
  && \left. \left. \mbox{} + 
  2\pi^2 \log \frac{\bar{M}^2}{m_i^2} \delta(x) \right]
  + m_i^2 K_0(m_ix) \dsl \frac{1}{x^2} \right\}  
  \, ,
\eea
where $K_0$ is a modified Bessel function ~\cite{Abram} and
$\bar{M}= 2M/ \gamma_E$, with $\gamma_E=1.781\dots$ the Euler's constant.
The scale dependence is
\be
  M \frac{\d}{\d M} <\Psi_i^A(x_1)\bar{\Psi}_i^B(x_2)> =
  \frac{1}{16\pi^2} g^2 \delta^{AB} \frac{N_c^2-1}{N_c}
  (\dsl-4m_i) \delta(x) \, .
\ee

The renormalized {\bf ghost selfenergy} is
\bea
  <\eta^a(x_1)\bar{\eta}^b(x_2)> &=& 
  -\frac{1}{2} g^2 N_c \delta^{ab} \Box \B^R[1] \nn
  &=&
  \frac{1}{128\pi^4} g^2 \delta^{ab} N_c \Box\Box 
  \frac{\log x^2 M^2}{x^2} \, ,
\eea
and its scale dependence,
\be
  M \frac{\d}{\d M} <\eta^a(x_1)\bar{\eta}^b(x_2) =
  - \frac{1}{16\pi^2} g^2 \delta^{ab} N_c \Box \delta(x) \, .
\ee

For the three-point functions we use the shifted variables 
$x=x_1-x_3$, $y=x_3-x_2$. Unless otherwise specified, all basic 
functions are 
assumed to depend on these two variables in the following. 
To avoid too lengthy expressions, we shall only give the 
final expressions in terms of basic functions.
The renormalized expression of the {\bf quark vertex} is
\bea
  \lefteqn{<\Psi_i^A(x_1)\bar{\Psi}_i^B(x_2)A_\mu^a(x_3)> = 
  - \ii g^3 (T^a)^{BA} 
  \left\{ \frac{1}{N_c} \left[
  \left(-2 m_i \d_\mu^+ + m_i^2 \gamma_\mu \right. \right. \right. } && \nn
  && \left. \mbox{} + 
  \dsl^x \gamma_\mu \dsl^y\right) \T_{m_im_i0}[1] +
  \left(4m_i\delta_{\mu\alpha}- \gamma_\alpha \gamma_\mu \dsl^y - 
  \dsl^x \gamma_\mu \gamma_\alpha\right) \T_{m_im_i0}[\d_\alpha] \nn
  && \left. \mbox{} - 
  \gamma_\mu \T^R_{m_im_i0}[\Box]  +
  2 \T^R_{m_im_i0}[\d_\mu \dsl] \right] + N_c \left[
  \frac{3}{2}m_i (\dsl^y\gamma_\mu+\gamma_\mu\dsl^x) \T_{00m_i}[1] 
  \right. \nn
  && \mbox{} + \left(\frac{3}{2} (\dsl^x\gamma_\mu\gamma_\alpha + 
  \gamma_\alpha\gamma_\mu\dsl^y) + \gamma_\mu \d_\alpha^+  
  -2 \gamma_\alpha \d_\mu^+ - \delta_{\mu\al} (3m_i +2 \dsl^+)  
  \right) \T_{00m_i}[\d_\al] \nn
  && \left. \left. \mbox{} + 
  \gamma_\mu \T^R_{00m_i}[\Box] + 2\T^R_{00m_i}[\d_\mu\dsl] 
  \rule[-.3cm]{0cm}{.6cm} \right] \right\} \, ,
\eea
where we have introduced the notation $\d^+ = \d^x+\d^y$. 
The scale dependent part reduces to
\be
  M \frac{\d}{\d M} <\Psi_i^A(x_1)\bar{\Psi}_i^B(x_2)A_\mu^a(x_3)> =
  \frac{\ii}{16\pi^2} g^3 (T^a)^{BA} \left(3N_c-\frac{1}{N_c}\right)
  \gamma_\mu \delta(x)\delta(y) \, .
\ee

For the {\bf ghost vertex} we have:
\bea
  \lefteqn{<\eta^b(x_1)\bar{\eta}^c(x_2)A_\mu^a(x_3)>  =  
  \frac{1}{2} g^3 N_c f^{abc} 
  \left\{\d^x\cdot\d^y\d_\mu^y \T[1] \right.} && \nn
  && \left. \mbox{} + \left(\delta_{\mu\alpha}(\Box^y-2\d^x\cdot\d^y)
  +\d_\mu^-\d_\alpha^y+\d_\mu^y\d_\alpha^-  \right) \T[\d_\alpha] 
  + \d_\mu^y \T^R[\Box] 
  \right\} \, ,
\eea
where $\d^- = \d^x-\d^y$ and $\d^x\cdot\d^y = \d^x_\alpha\d^y_\alpha$.
The scale dependence is
\be
  M \frac{\d}{\d M} <\eta^b(x_1)\bar{\eta}^c(x_2)A_\mu^a(x_3)> =
  \frac{1}{16\pi^2} N_c f^{abc} \delta(x_1-x_3) 
  \d_\mu^{x_2}\delta(x_2-x_3) \, .
\ee
Here, we have come back to the original variables, to make explicit that it 
has the same form as the corresponding term in the Lagrangian.

The renormalized expression of the {\bf triple gluon vertex} 
is quite large, even in terms of basic functions. We split 
it into the pure gauge part and the fermionic part. The results are 
\bea
  \lefteqn{<A_\mu^a(x_1)A_\nu^b(x_2)A_\rho^c(x_3)>^{G} = g^3 f^{abc} N_c
  } && \nn
  && \mbox{} \times \left\{ \rule[-.4cm]{0cm}{.8cm} \left[
  \frac{9}{4} \delta_{\nu\rho}\d_\mu^- \left(\B[1](x)\delta(x+y) \right) 
  + \frac{9}{4} (\delta_{\mu\nu}\d_\rho^x - \delta_{\mu\rho}\d_\nu^x) 
  \B[1](x)\delta(y) \right. \right.  \nn
  && \mbox{} +
  \frac{1}{2} \left( 3\d_\rho^y\d_\mu^x\d_\nu^- 
  + \delta_{\mu\nu} (\Box^x + 3 \d^x\cdot\d^y) \d_\rho^y  +
  \delta_{\mu\rho} (5\Box^x \d_\nu^y - 2(\Box^y+2\d^x\cdot\d^y)\d_\nu^x)
  \right) \T[1] \nn
  && \mbox{} +
  \frac{1}{2} \left( 
  \delta_{\mu\nu} (\d_\rho^x\d_\al^y - 3\d_\rho^x\d_\al^x) +
  \delta_{\mu\rho} (8\d_\nu^x\d_\al^x-4\d_\nu^y\d_\al^x -3\d_\nu^y\d_\al^y) 
  \right. \nn
  && \mbox{} + 
  \delta_{\rho\al} \left(
  8\d_\mu^y\d_\nu^x-2\d_\mu^x\d_\nu^y-3\d_\mu^x\d_\nu^x
  +\delta_{\mu\nu} (\Box^x-7\d^x\cdot\d^y) \right)  \nn
  && \mbox{} +
  \left. \delta_{\mu\al} \left(
  6\d_\nu^y\d_\rho^y-13\d_\nu^x\d_\rho^y-\d_\nu^y\d_\rho^x-3\d_\nu^x\d_\rho^x
  + \delta_{\nu\rho} (\Box^x-10\Box^y+10\d^x\cdot\d^y) \right)
  \right) \T[\d_\al] \nn
  && \mbox{}  +
  \frac{1}{2} \left(2\delta_{\mu\nu}\d_\rho^x + 
  \delta_{\mu\rho}(2\d_\nu^y-5\d_\nu^x)\right) \T^R[\Box]
  + 4 \d_\rho^x \T^R[\d_\mu\d_\nu] + 4\d_\nu^y \T^R[\d_\mu\d_\rho] \nn
  && \mbox{} + \left. 
  \delta_{\mu\nu} \d_\alpha^x \T^R[\d_\al\d_\rho] +
  \delta_{\mu\rho}\d_\al^y \T^R[\d_\al\d_\nu] -
  4 \T^R[\d_\mu\d_\nu\d_\rho] \right] \nn
  && \mbox{} + \left.
  \left[ \begin{array}{c}
  x \leftrightarrow y \\
  \mu \leftrightarrow \nu 
  \end{array} \right] \right\} 
\eea
for the pure gauge and
\bea
  \lefteqn{<A_\mu^a(x_1)A_\nu^b(x_2)A_\rho^c(x_3)>^{F} = 2 g^3 f^{abc}
  \sum_{i=1}^{N_f} \left\{ \rule[-.4cm]{0cm}{.8cm} \left[
  m_i^2 \left(\delta_{\mu\nu}\d_\rho^x - \delta_{\mu\rho}
  \d_\nu^-\right) \T_{m_im_im_i}[1] \right. \right. } && \nn
  && \mbox{} +
  \left( \delta_{\mu\nu}\d_\rho^x\d_\al^y
  - \delta_{\mu\rho} (\d_\nu^x\d_\al^y-\d_\nu^y\d_\al^x)
  + \frac{1}{2} \delta_{\rho\al} \left(\d_\mu^y\d_\nu^x-\d_\mu^x\d_\nu^y-
  \delta_{\mu\nu}(\d^x\cdot\d^y+m_i^2) \right) \right. \nn
  && \left. \mbox{} +
  \delta_{\mu\al} \left(-\d^x_\nu\d_\rho^y-\d_\nu^y\d_\rho^x
  +\delta_{\nu\rho} (\d^x\cdot\d^y-m_i^2) \right) 
  \right) \T_{m_im_im_i}[\d_\al] \nn
  && \mbox{} -
  \left(\delta_{\mu\nu} \d_\rho^x - \delta_{\mu\rho}\d_\nu^- \right)
  \T_{m_im_im_i}^R[\Box]
  + 2 \d_\rho^x \T_{m_im_im_i}^R[\d_\mu\d_\nu]
  + 2 \d_\nu^y \T_{m_im_im_i}^R[\d_\mu\d_\rho] \nn
  && \mbox{} -  
  2 \delta_{\mu\rho}\d_\al^x \T_{m_im_im_i}^R[\d_\al\d_\nu]
  + \frac{1}{2} \delta_{\mu\nu} \T_{m_im_im_i}^R[\Box\d_\rho] 
  + \delta_{\mu\rho}  \T_{m_im_im_i}^R[\Box\d_\nu] \nn
  && \mbox{} - \left.
  2 \T_{m_im_im_i}^R[\d_\mu\d_\nu\d_\rho] \right] + 
  \left. \left[ \begin{array}{c}
  x \leftrightarrow y \\
  \mu \leftrightarrow \nu 
  \end{array} \right] \right\} 
\eea
for the fermionic part.
The Bose symmetry of the three gluons is not obvious because
of the use of the shifted variables. Only the symmetry under interchange of
$A_\mu^a(x_1)$ and $A_\nu^b(x_2)$ (corresponding to
$(x,y,a,\mu) \leftrightarrow (-y,-x,b,\nu)$) is explicit. 
The complete scale dependence, written
in the original variables, reads
\bea
  \lefteqn{M\frac{\d}{\d M} <A_\mu^a(x_1)A_\nu^b(x_2)A_\rho^c(x_3)> =   
  \frac{1}{12\pi^2} g^3 f^{abc} (N_c-N_f)} && \nn
  && \mbox{} \times
  \left[ \delta_{\mu\nu} (\d_\rho^{x_1}-\d_\rho^{x_2}) +
  \delta_{\nu\rho}(\d_\mu^{x_2}-\d_\mu^{x_3}) +
  \delta_{\rho\mu} (\d_\nu^{x_3}-\d_\nu^{x_1})\right]
  \left[ \delta(x_1-x_3)\delta(x_2-x_3) \right] \, .
\eea

In Ref.~\cite{FGJR}, the bare triple gluon vertex in the 
background field
formalism (with massless quarks) was found to be conformal invariant 
for non-coincident points, if the Feynman gauge is employed. 
This fact, together with the fulfilment of the Ward identities, 
implies that this three-point function must be a linear combination of
the two permutation odd conformal tensors
$D_{\mu\nu\rho}^{\mbox{\footnotesize symm}}(x_1,x_2,x_3)$ 
and $C_{\mu\nu\rho}^{\mbox{\footnotesize symm}}(x_1,x_2,x_3)$~\cite{Schreier}.
The explicit expression in terms of these conformal tensors was also found in 
Ref.~\cite{FGJR}. Differential
renormalization was then used to treat the singularities at coincident points.
Of course, renormalization breaks conformal invariance, but the Ward identity
relating the triple background gluon vertex to the background gluon
selfenergy was enforced by adequately adjusting the
renormalization scales that appear in the process. We have applied CDR
to the renormalization of the conformal tensors 
$D^{\mbox{\footnotesize symm}}$ and $C^{\mbox{\footnotesize symm}}$, 
and checked that the resulting amplitude directly fulfils the Ward identity
(if the CDR result for the gluon selfenergy
is used)\footnote{It is important that we have used the form of the
conformal tensors given by Eqs. (4.1) and (4.16) of Ref.~\cite{FGJR}.
The triple gluon vertex can be expressed in terms of the
conformal tensors in this form using just the Feynman rules and 
the Leibniz rule for derivatives, which is an allowed operation in CDR.
Hence, no ambiguity is introduced in the process previous to our calculation.}. 
No adjustment is needed {\it a posteriori}. The tensor 
$C^{\mbox{\footnotesize symm}}$ is finite, but ambiguous, and the result in 
CDR differs from the
one given in Ref.~\cite{FGJR} by a finite local term.
In our case, it gives a non-vanishing
contribution to the Ward identity. The result for 
$D^{\mbox{\footnotesize symm}}$ also has an extra local term with respect to
the final one in Ref.~\cite{FGJR}.
The discrepancies are due to the fact that, while in CDR everything is
fixed from the start, in conventional differential renormalization 
the renormalization scales can be adjusted, in general,  
in more than one manner to preserve the Ward identities.

Finally, we have calculated the {\bf quartic gluon vertex}, 
but the final expression is too
lengthy, even in terms of basic functions, to be written here. 
We only give the scale dependent part:
\bea
  \lefteqn{M\frac{\d}{\d M} <A_\mu^a(x_1)A_\nu^b(x_2)A_\rho^c(x_3)
  A_\sigma^d(x_4)> = } && \nn
  && \mbox{} 
  -\frac{1}{24\pi^2} g^4 (N_c+2N_f)  
  \left[f^{abr}f^{cdr} (\delta_{\mu\rho}\delta_{\nu\sigma}-
  \delta_{\mu\sigma}\delta_{\nu\rho})\right. \nn
  && \left. \mbox{} +
  f^{acr}f^{dbr} (\delta_{\mu\sigma}\delta_{\nu\rho}-
  \delta_{\mu\nu}\delta_{\rho\sigma}) +
  f^{adr}f^{bcr} (\delta_{\mu\nu}\delta_{\rho\sigma}-
  \delta_{\mu\rho}\delta_{\nu\sigma}) \right] \, .
\eea

The renormalized 1PI Green functions we have calculated satisfy
the renormalization group equation
\be
  \left[M \frac{\d}{\d M} + \beta \frac{\d}{\d g} + \gamma_{m_i} m_i 
  \frac{\d}{\d m_i}
  - n_A \gamma_A - n_\eta \gamma_\eta - 
  \sum_{i=1}^{N_f} n_{\Psi_i} \gamma_{\Psi_i}  \right]
  \Gamma^{(n_\A,n_\eta,n_{\Psi_1},\dots)} = 0 \, ,
\ee
where $n_\A$, $n_\eta$ and $n_{\Psi_i}$ are the number of gauge, ghost and 
$i$-quark fields, respectively. The coefficients can be easily obtained from the
scale dependent parts given above; the standard values are recovered:
\bea
  \beta & = & - \frac{g^3}{48\pi^2} (11N_c-2N_f) \, , \\
  \gamma_A & = & -\frac{g^2}{48\pi^2} (5N_c-2N_f) \, , \\
  \gamma_\eta & = & - \frac{g^2}{32\pi^2} N_c \, , \\
  \gamma_{\Psi_i} & = & \frac{g^2}{32\pi^2} \frac{N_c^2-1}{N_c} \, , \\
  \gamma_{m_i} & = & - \frac{3}{16\pi^2} g^2 \frac{N_c^2-1}{N_c} \, .
\eea
Note that $N_c$ and $\frac{N_c^2-1}{2N_c}$ are the $SU(N)$ 
Casimir invariants of the adjoint and fundamental representation, respectively.
Unlike in the background field method, the anomalous dimension of
the gauge field is not directly related to the $\beta$ function.
The same $\beta$ function is obtained from all vertex functions,
showing that there is a single coupling $g$. This is a consequence of
the Slavnov-Taylor identities for the scale dependent parts of the Green
functions. 

Let us now write the full set of Slavnov-Taylor identities for the complete 
1PI Green functions. They are a bit more involved than the ones for
connected Green funtions \cite{connectedST}. 
The general form in terms of the effective action
can be directly derived from the BRST symmetry of the Lagrangian
\cite{ZJL}. Using the source terms given by Eq.~(\ref{sources}) and suppressing 
all indices it reads
\be
  \int {\mathrm d}^4x \, \left[
  \frac{\delta\Gamma}{\delta A} \frac{\delta\Gamma}{\delta K} -
  \frac{\delta\Gamma}{\delta \eta} \frac{\delta\Gamma}{\delta L} + 
  \frac{\delta\Gamma}{\delta N} \frac{\delta\Gamma}{\delta \bar{\Psi}} -
  \frac{\delta\Gamma}{\delta \Psi} \frac{\delta\Gamma}{\delta \bar{N}}
  + \frac{\delta\Gamma}{\delta \bar{\eta}} \d A \right] = 0 \, .
\label{genST}
\ee
To one loop, the quadratic terms can be ``linearized'':
\be
  \frac{\delta\Gamma}{\delta \phi} \frac{\delta\Gamma}{\delta J_{s\phi}} =
  \frac{\delta\Gamma^{(0)}}{\delta \phi} 
  \frac{\delta\Gamma^{(1)}}{\delta J_{s\phi}} +
  \frac{\delta\Gamma^{(1)}}{\delta \phi} 
  \frac{\delta\Gamma^{(0)}}{\delta J_{s\phi}} \, .
\ee
On the other hand, the ghost equation of motion,
\be
  \d_\mu \frac{\delta\Gamma}{\delta K_\mu^a} + 
  \frac{\delta\Gamma}{\delta \bar{\eta}^a} = 0 \, ,
\ee
allows to simplify the identity for the gluon selfenergy.
This equation is trivially fulfilled to all orders in any renormalization
scheme commuting with differentiation and preserving the structure of the 
Lagrangian. 
\begin{figure}[ht]
\begin{center}
\epsfxsize=12cm
\epsfbox{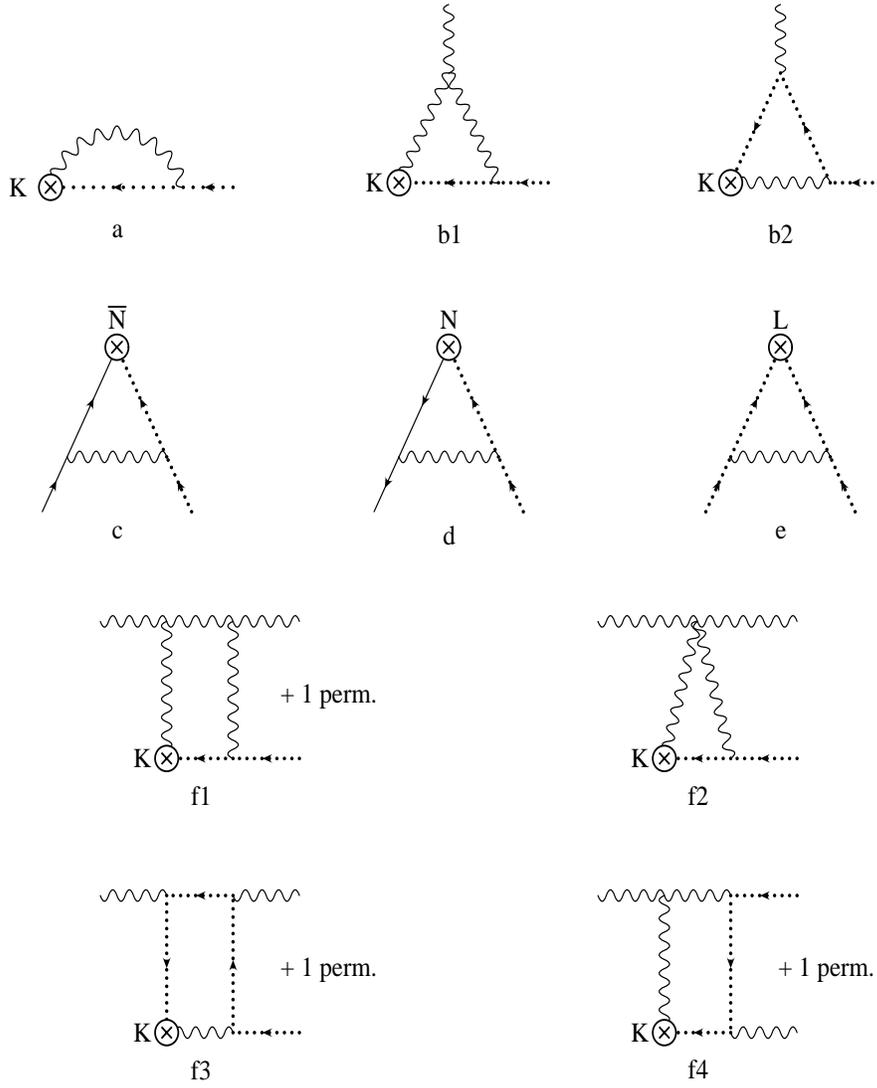}
\end{center}
\caption{One-loop Feynman diagrams contributing to 1PI Green functions
with insertions of BRST variations. The Feynman diagrams for singular 1PI Green
functions of elementary fields can be found in Ref.~\cite{PT}.
The diagrams contributing 
to the (finite) gluon-gluon-ghost-antighost function, which appears in the
quartic-gluon-vertex identity, are identical to diagrams f1-f4, but changing
$K$ by an antighost external leg. \label{fig_insertions}}
\end{figure}
Writing explicitly the tree-level pieces, the one-loop 
Slavnov-Taylor identities read (we name each identity 
after the Green function with the largest number of elementary fields)
\begin{itemize}
\item {\em gluon selfenergy identity:}
\be
  0 = \d_\mu^x <\A_\mu^a(x) A_\nu^b(y)>  \, , 
\ee
\item {\em quark vertex identity:}
\bea
  \lefteqn{0=g \gamma_\mu (T^a)^{AB} \delta(x-y) <\eta^b(z)K_\mu^a(x)> -
  \d_\mu^z <\Psi_i^B(x)\bar{\Psi}_i^A(y)A_\mu^b(z)>}&&  \nn
  && \mbox{} -(\dsl^y+m_i) <\Psi_i^B(x)\eta^b(z)\bar{N}_i^A(y)> 
  - <\eta^b(z)\bar{\Psi}_i^A(y)N_i^B(x)> 
  (\stackrel{\leftarrow}{\dsl^x}-m_i)  \nn
  && \mbox{} - g (T^b)^{CB} \delta(x-z) <\Psi_i^C(x)\bar{\Psi}_i^A(y)> \nn
  && \mbox{} + g (T^b)^{AC} \delta(y-z) <\Psi_i^B(x) \bar{\Psi}_i^C(y)> \, , 
\eea
\item {\em ghost vertex identity:}
\bea
  \lefteqn{0=\left[\d_\mu^x <A_\mu^b(x) \eta^c(y)\bar{\eta}^d(z)>   
  + g f^{acd}\d_\mu^z\left(\delta(y-z)<\eta^b(x)K_\mu^a(z)> \right) 
  \right]} && \nn
  &&\mbox{} - \left[\begin{array}{c}
  b\leftrightarrow c \\
  x\leftrightarrow y \end{array} \right] 
  + \Box^z <\eta^b(x)\eta^c(y)L^d(z)> 
  + g f^{abc}\delta(x-y) <\eta^a(x)\bar{\eta}^d(z)> \, , 
\eea
\item {\em triple gluon vertex identity:}
\bea
  \lefteqn{0= \left[
  g f^{adb} \delta(y-z) <A_\nu^c(x)A_\rho^a(y)> -  
  \Box^x <\eta^b(z) A_\rho^d(y) K_\nu^c(x)> \right.}&& \nn
  && \mbox{} \left.
  - \d_\nu^x <\eta^b(z)\bar{\eta}^c(x)A_\rho^d(y)> \right] +
  \left[\begin{array}{c}
  c\leftrightarrow d \\
  \nu\leftrightarrow \rho \\
  x\leftrightarrow y 
  \end{array} \right] \nn
  && \mbox{} + g f^{acd} \left(-\delta_{\mu\nu}
  (2\d_\rho^x+\d_\rho^y)+\delta_{\nu\rho}(\d_\mu^x-\d_\mu^y)+
  \delta_{\rho\mu}(\d_\nu^x+2\d_\nu^y)\right) \nn
  && 
  \left(\delta(x-y) <\eta^b(z)K_\mu^a(x)>\right)
  -\d_\mu^z <A_\mu^b(z)A_\nu^c(x)A_\rho^d(y)>\, ,
\eea
\item {\em quartic gluon vertex identity:}
\bea
  \lefteqn{0= \left[
  g f^{acd} \left(-\delta_{\mu\nu}(2\d_\rho^y+\d_\rho^z) + \delta_{\nu\rho}
  (\d_\mu^y-\d_\mu^z) + \delta_{\rho\mu}(\d_\nu^y+2\d_\nu^z)\right) \right.}
  && \nn
  && \mbox{} \left(\delta(y-z) <\eta^b(t)A_\sigma^e(x)K_\mu^a(y)> \right)
  + g f^{bae}\delta(x-t) <A_\nu^c(y)A_\rho^d(z) A_\sigma^a(t)> \nn
  && \left.\mbox{} - \Box^x <A_\nu^c(y)A_\rho^d(z)\eta^b(t)K_\sigma^e(x)>
  - \d_\sigma^x <A_\nu^c(y)A_\rho^d(z) \eta^b(t)\bar{\eta}^e(x)> \right]
  \nn && \mbox{} + \left[\begin{array}{c}
  e\leftrightarrow c \\
  \sigma\leftrightarrow \nu \\
  x\leftrightarrow y
  \end{array} \right]
  + \left[\begin{array}{c}
  e\leftrightarrow d \\
  \sigma\leftrightarrow \rho \\
  x\leftrightarrow z 
  \end{array} \right] \nn
  && \mbox{} + 
  g^2 \left(f^{acr}f^{der}(\delta_{\mu\rho}\delta_{\nu\sigma}
  -\delta_{\mu\sigma}\delta_{\nu\rho}) + f^{adr}f^{ecr}
  (\delta_{\mu\sigma}\delta_{\nu\rho}-\delta_{\mu\nu}\delta_{\rho\sigma})
  \right. \nn
  && \left. \mbox{} + f^{aer}f^{cdr} (\delta_{\mu\nu}\delta_{\rho\sigma}-
  \delta_{\mu\rho}\delta_{\nu\sigma})\right)  
  \left(\delta(x-y)\delta(x-z) <\eta^b(t)K_\mu^a(x)>\right) \nn
  && \mbox{} -\d_\mu^t <A_\mu^b(t)A_\nu^c(y)A_\rho^d(z)A_\sigma^e(x)> \, .
\eea
\end{itemize}
The diagrams contributing to Green functions with insertions of BRST
variations are depicted in Fig.~\ref{fig_insertions}.
Verifying that the renormalized Green functions calculated above 
satisfy these identities is not straightforward, due
to the length of the expressions involved and to the fact that we are
not dealing with simple functions but with distributions. The obvious
way to compare distributions is to make them act on a general test function. 
In particular one can perform a Fourier transform without loss of 
information\footnote{The necessary Fourier transforms are collected in 
Appendix~B of Ref.~\cite{techniques}.}. 
This has the advantage that the resulting finite integrals can be treated with
standard momentum space techniques and that, for 
non-exceptional momenta, no infrared divergencies appear. 
We have used a {\em Mathematica}-based
program to carry out the algebraic operations, linked to {\em
LoopTools} \cite{HPV}, which calculates numerically one-loop integrals.
As advanced, CDR respects the Slavnov-Taylor identities. 

Summarizing, we have applied CDR to the one-loop singular 1PI 
Green functions of QCD and have verified that the Slavnov-Taylor identities 
are preserved. The extension to Yang-Mills theories with a more general 
gauge group does not introduce new complications, as far as renormalization 
is concerned, so CDR should treat the general case equally well. Here we have 
sticked to $SU(N_c)$ because the computer implementation is 
simpler~\cite{Verm}. 
One could also wonder about the performance of CDR when the gauge symmetry is
spontanously broken. Again, as CDR does not depend on the structure 
of the interaction Lagrangian, there should be no extra problems in dealing 
with this case. The authors of Ref.~\cite{HPV}, using the programs described 
there, have checked that CDR renders a transverse vacuum
polarization in the electroweak standard model, and recovers the standard 
physical results for Z-Z and W-W elastic scattering. These examples also
test the inclusion of scalar fields in a non-Abelian gauge theory.
Finally, let us point out that 
very recently it has been found (in momentum space) that CDR and regularization
by dimensional reduction produce equivalent results at the one loop level 
\cite{HPV}. This
gives an alternative explanation for the preservation of gauge invariance in
CDR. 

\section*{Acknowledgments}
I thank F. del Aguila for discussions. Many of the calculations have been
performed with computers of the Institut f\"ur Teoretische Physik
of the University of Karlsruhe. This work has been supported by CICYT, under
contract number AEN96-1672 and by Junta de Andaluc\'{\i}a, FQM101.
I also thank Ministerio de Educaci\'on y Cultura for financial support.


\begin{thebibliography}{99}
\bibitem{FJL} D.Z. Freedman, K. Johnson and J.I. Latorre, 
    \np{B371}{92}{353}.
\bibitem{CDR} F. del Aguila, A. Culatti, R. Mu\~noz Tapia and
    M. P\'erez-Victoria, \pl{B419}{98}{263}; 
    F. del Aguila and M. P\'erez-Victoria, 
    \vj{Acta Phys. Polon.}{B28}{97}{2279}.
\bibitem{techniques} F. del Aguila, A. Culatti, R. Mu\~noz Tapia and
    M. P\'erez-Victoria, MIT-CTP-2705, UG-FT-86/98, 
    hep-ph/9806451.
\bibitem{g2} F. del Aguila, A. Culatti, R. Mu\~noz Tapia and
    M. P\'erez-Victoria, \np{B504}{97}{532}; F. del Aguila, A. Culatti, R. Mu\~noz Tapia and
    M. P\'erez-Victoria, International Workshop on
    Quantum Effects in MSSM, Universitat Aut\`onoma de
    Barcelona, September 1997, hep-ph/9711474.
\bibitem{ST} A.A. Slavnov, \vj{Theor. Math. Phys.}{10}{72}{152} (English
    translation: \vj{Theor. and Math. Phys.}{10}{72}{99});
    J.C.Taylor, \np{B33}{71}{436}.
\bibitem{FGJR} D.Z. Freedman, G. Grignani, K. Johnson and N. Rius,
    \ap{218}{92}{75}.
\bibitem{BGM} B.S. DeWitt, \pr{162}{67}{1195,1239};
    G.~'t~Hooft, \np{B62}{73}{444};
    L.F. Abbott, \np{B185}{81}{189}.
\bibitem{Diagrammar} G. 't Hooft and M. Veltman,
    {\em Diagrammar}, CERN 73-9 (1973).
\bibitem{ZJL} J. Zinn-Justin, in {\em Trends in Elementary Particle Theory,
    International Summer Institute on Theoretical Physics, Bonn 1974}
    (Springer-Verlag, Berlin, 1975); B.W. Lee, in {em Methods in Field Theory,
    Les Houches 1975}, eds. R. Balian and J. Zinn-Justin (North-Holland,
    Amsterdam).
\bibitem{BRST} C. Becchi, A. Rouet and R. Stora, \vj{Comm. Math.
    Phys.}{42}{75}{127}; \ap{98}{76}{287};
    I.V. Tyutin, Lebedev Institute preprint N39 (1975).
\bibitem{PT} P. Pascual and R. Tarrach, {\em QCD: Renormalization for the
    Practitioner} (Springer-Verlag, Berlin, Heidelberg, New York, Tokyo,
    1984).
\bibitem{Abram} M. Abramowitz and I.A. Stegun,
    {\em Handbook of Mathematical Functions} (Dover, New York, 1972).
\bibitem{Schreier} E.J. Schreier, \pr{D3}{71}{980}.
\bibitem{connectedST} E.S. Abers and B.W. Lee, \vj{Phys. Rep.}{9C}{73}{1};
     W. Marciano and H. Pagels, \vj{Phys. Rep.}{36C}{78}{137}.
\bibitem{HPV} T. Hahn and M. P\'erez-Victoria, UG-FT-87/98, KA-TP-7-1998,
     hep-ph/9807565.
\bibitem{Verm} J.A.M. Vermaseren, {\em The use of computer algebra in QCD}, in H.
     Latal, W. Schweiger, Proceedings Schladming 1996, (Springer ISBN
     3-540-62478-3); T. van Ritbergen, A.N. Schellekens and J.A.M. Vermaseren,
     NIKHEF-98-004,UM-TH-98-01, hep-ph 9802376.
\end{thebibliography}
\end{document}